\documentclass[aps,prb,preprint,superscriptaddress,floatfix]{revtex4-1}

\usepackage{graphicx}
\usepackage{dcolumn}
\usepackage{bm}
\usepackage{upgreek}
\usepackage{natmove}

\usepackage{xcolor}

\begin{document}


\title{Surface flaws control strain localization in the deformation of Cu$\vert$Au nanolaminates}


\author{Adrien Gola}
\affiliation{Department of Microsystems Engineering, University of Freiburg, Georges-K\"ohler-Allee 103, 79110 Freiburg, Germany}
\affiliation{Institute for Applied Materials, Karlsruhe Institute of Technology (KIT), Hermann-von-Helmholtz-Platz 1, 76344 Eggenstein-Leopoldshafen, Germany}

\author{Guang-Ping Zhang}
\affiliation{Shenyang National Laboratory for Materials Science,
	Institute of Metal Research, Chinese Academy of Sciences, 72 Wenhua Road, Shenyang 110016, P.R. China}

\author{Lars Pastewka}
\email{lars.pastewka@imtek.uni-freiburg.de}
\affiliation{Department of Microsystems Engineering, University of Freiburg, Georges-K\"ohler-Allee 103, 79110 Freiburg, Germany}
\affiliation{Institute for Applied Materials, Karlsruhe Institute of Technology (KIT), Hermann-von-Helmholtz-Platz 1, 76344 Eggenstein-Leopoldshafen, Germany}
\affiliation{Freiburg Materials Research Center, University of Freiburg, 79104 Freiburg, Germany}
\affiliation{Cluster of Excellence livMatS @ FIT -- Freiburg Center for Interactive Materials and Bioinspired Technologies, University of Freiburg, Georges-Köhler-Allee 105, 79110 Freiburg, Germany}

\author{Ruth Schwaiger}
\affiliation{Institute for Applied Materials, Karlsruhe Institute of Technology (KIT), Hermann-von-Helmholtz-Platz 1, 76344 Eggenstein-Leopoldshafen, Germany}



\date{\today}

\begin{abstract}
We carried out matched experiments and molecular dynamics simulations of the compression of nanopillars prepared from Cu$\vert$Au nanolaminates with $25$~nm layer thickness. 
The stress-strain behavior obtained from both techniques are in excellent agreement. 
Variation of the layer thickness in simulations reveals an increase of the strength with decreasing layer thickness.
Pillars fail through the formation of shear bands whose nucleation we trace back to the existence of surface flaws. Our combined approach demonstrates the crucial role of contact geometry in controlling the deformation mode and suggests that modulus-matched nanolaminates should be able to suppress strain localization while maintaining controllable strength. 
\end{abstract}


\maketitle


Mechanical properties of materials deviate from bulk behavior when characteristic dimensions become small. Such deviations may occur when either microstructural features, e.g. the grain size, or object dimensions, approach the length scale of the process that controls the deformation. As a result, the mechanical strength of micro- or nano-scale pure metallic materials has been found to be an order of magnitude higher than their bulk counterparts~\cite{ramesh2009nanomaterials,Uchic:2004p986,Greer:2011p654}.
A special class of nanostructured materials are metallic nanolaminates with nanoscale layers of two different materials. They not only exhibit enhanced strength and hardness~\cite{tsakalakos:1986,lehoczky:1978,misra:1998:nanolayers,misra2001,wang2011}, wear resistance~\cite{ruff:1991:wear,wen:2008:wear} or toughness~\cite{zhang:2010:cucr}, but also offer the possibility to tailor those properties by choosing material combinations~\cite{Gola:2018}.

Nanolaminates exhibit a range of different deformation behaviors, which depend on the combination of materials, type of interfaces~\cite{Zhang2014-fv} and thickness of the laminate layers~\cite{wang2011}. Reducing the thickness $\lambda$ of the layer increases the flow strength $\sigma$ of the material, with Hall-Petch-like behavior, $\sigma\propto \lambda^{-1/2}$ at large thickness transitioning to confined layer slip $\sigma\propto \ln(\lambda)/\lambda$ at smaller thickness. Shear band instabilities were observed for several crystalline systems and attributed to a reduced strain hardening ability~\cite{Wei:2002:1240,Wang:2011:7290}.
 Since shear-banding is the primary failure mechanism in nanolaminates under compression~\cite{Li:2009:728}, engineering a strong nanolaminate requires control or elimination of shear-banding.

The work presented here extends on the previous investigations in two important directions: First, mechanical tests are carried out by compressing micropillars rather than through indentation. Results of pillar compression tests are easier to interpret because in contrast to indentation testing, the stress experienced by the pillar is largely uniform and \emph{in situ} observation of pillars allows direct measurement of the deformation. Second, we present a first quantitative comparison between experiment and accompanying molecular dynamics (MD) simulations, the latter carried out on nanolaminates models at realistic scales and with realistic microstructures and boundary conditions~\cite{schwaiger:2012:jmr}. Simulations yield both mechanical properties as well as failure behavior of the pillars that can be directly compared with our experiments.
We specifically focus on the Cu$\vert$Au nanolaminate system that has been studied extensively over the past few years~\cite{Li:2010:3049,Luo:2015:67}. Cu$\vert$Au nanolaminates have a semi-coherent interface with a network of dislocations reducing the coherency stress in the layers~\cite{Gola2018-wb,Shao:2015:242}.

\begin{figure}
\centering
    \includegraphics[width=8.5cm,keepaspectratio]{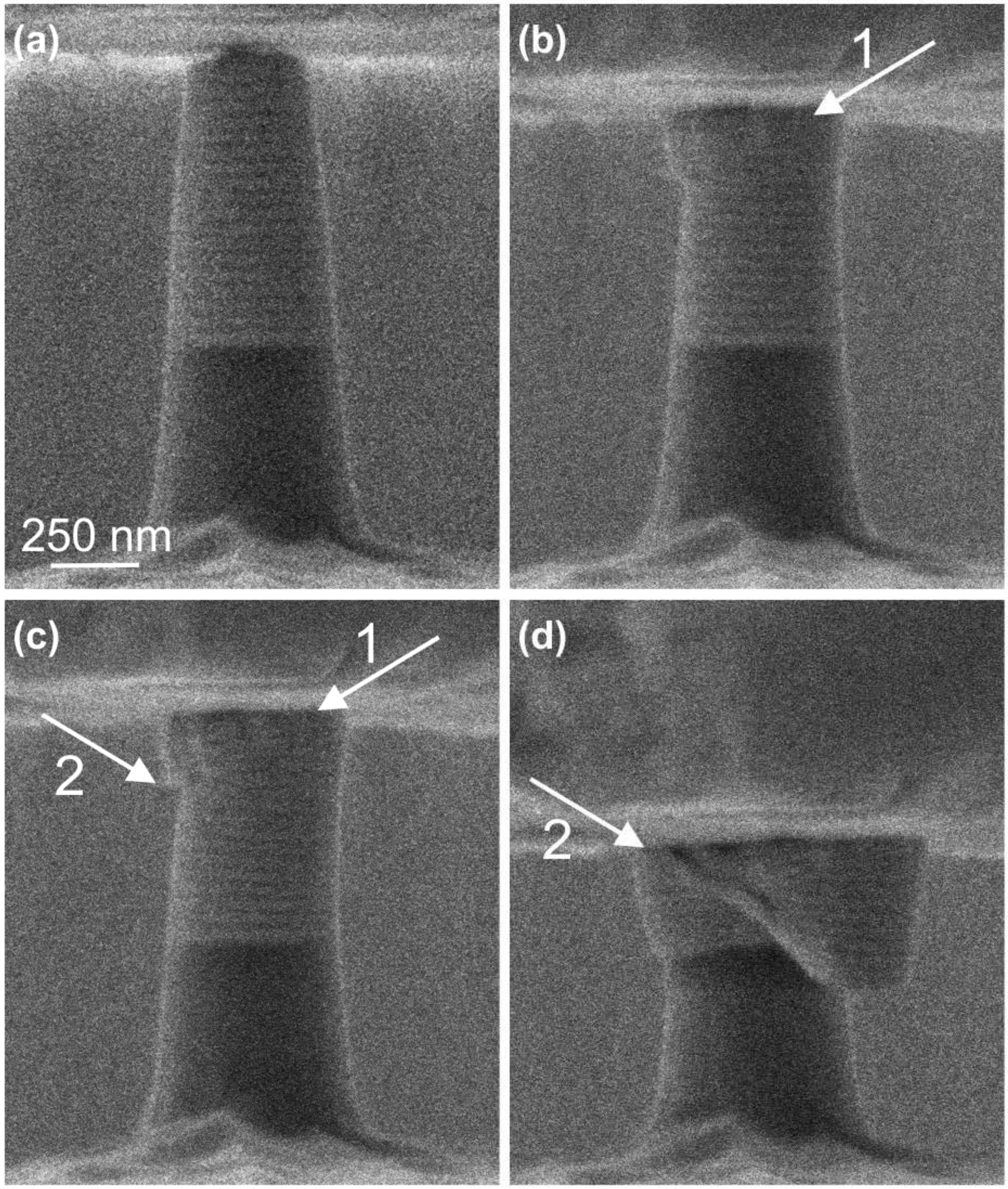}
    \caption{Scanning electron microscope images of deformation and failure during compression of a nanolaminate pillar comprised of $40$ layers of $\lambda = 25$~nm thickness. (a) Prior to deformation. (b) The deformation localizes at the top of the pillar and (c) forms a step on the top half (d) followed by failure through an interlayer shear-band.}
    \label{fig:sem_exp}
\end{figure} 

Our experimental nanopillars were prepared by focused ion beam (FIB) milling from a Cu$\vert$Au nanolaminate, which had a strong Cu and Au $\{111\}$ texture and had been sputter-deposited on a $(100)$ Si substrate~\cite{Zhang:2006:13105}. The pillars had a taper angle of 4$^{\circ}$ and diameters at the surface and at the interface of 370~nm and 480~nm, respectively. The actual test volume was composed of a $40$ layer stack of $25$~nm individual thickness giving a total sample thickness of $1~\upmu$m (Fig.~\ref{fig:sem_exp}a). The nanopillars were compressed \emph{in situ} in a scanning electron microscope (SEM, FEI Nova NanoLab 200 and Nanomechanics InSem nanoindenter) to observe their behavior during deformation.

Figure~\ref{fig:sem_exp}a-d shows the typical outcome of such an experiment. The diamond punch first contacted the pillar on its flat top (Fig.~\ref{fig:sem_exp}a). Deformation then led to the gradual compression of the pillar and eventually to the nucleation of a shear band (indicated by a ``1'' in Fig.~\ref{fig:sem_exp}b and c). Shear banding localized further deformation and led to the extrusion of a wedge-shaped region near the top of the pillar. Further compression  nucleated a second shear band, initiated right where the wedge had slid enough to create a surface step that concentrated stress (position ``2'' in Fig.~\ref{fig:sem_exp}c and d). Deformation then continued along this secondary shear band and eventually resulted in the failure of the pillar.

The experimental observations pose two important questions: First, it is unclear which process sets the strength of the material and which role the layer thickness plays in that process. We note that there is no evidence for slip along the interface in these pillars. Experiments on pillars with tilted interfaces and simulations on representative volume elements suggest that the interfacial shear strength is $\sim 0.3$~GPa (see Supplementary Section S-I) but the Schmid factor for sliding along the interface for the loading geometry shown in Fig.~\ref{fig:sem_exp} is zero. Second, homogeneous deformation was followed by the traversal of a shear band that led to the failure of the pillar. From the experiments alone, it remains unclear what conditions led to the nucleation of these shear bands. Experimental pillars often have defects from growth and FIB preparation, as for example surface roughness. We here hypothesize a primary reason must be symmetry breaking due to the existence of surface flaws on either pillar or indenter tip.

\begin{figure}
\centering
    \includegraphics[width=8.5cm,keepaspectratio]{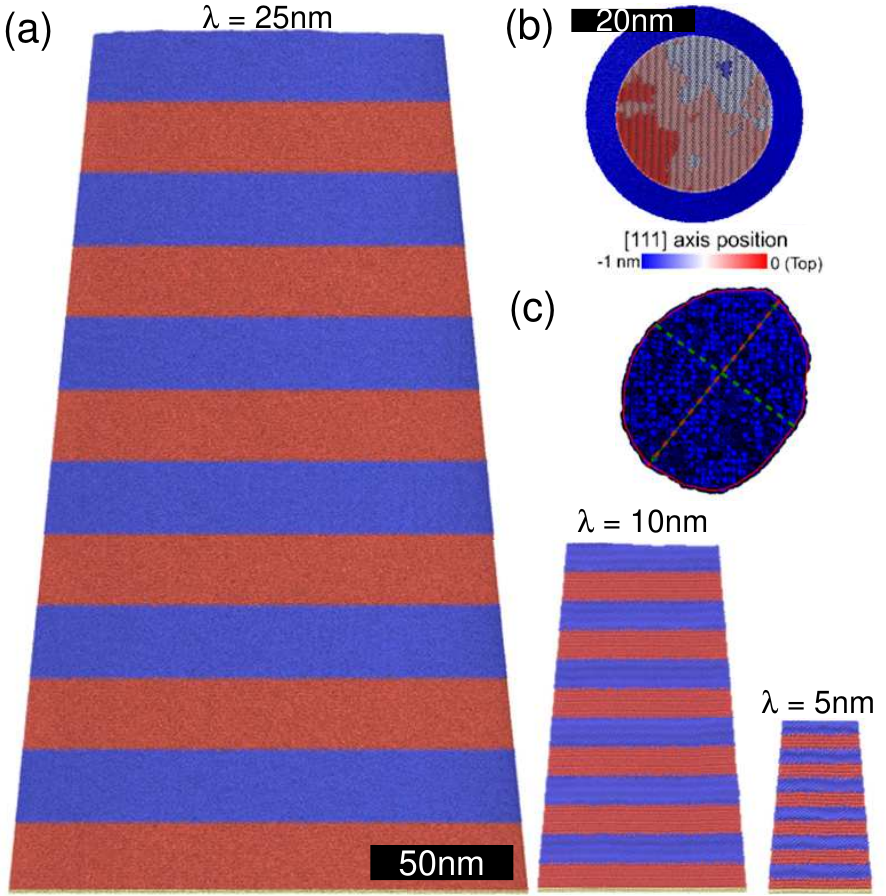}
    \caption{(a) Side view of our atomistic pillar models with layer thickness of $\lambda = 25$, $10$ and $5$~nm. Atoms are color coded according to their type with Cu atoms in blue and Au atoms in red. The yellow atoms at the bottom are a rigid substrate of Au atoms. (b) Top view of the pillar model with $\lambda = 5$~nm showing a realization of random surface roughness. Atoms are colored after their position along the $[111]$ crystallographic direction that is normal to the interfaces. (c) Cross section at $1/5$ of the pillar height during compression used to compute the cross sectional area from the MD calculations. Red and green dashed lines show the longest and shortest half-axes of the cross section.}
    \label{fig:setup}
\end{figure}

To test this hypothesis, we carried out molecular dynamics (MD) calculations with varying layer thickness from $5$~nm to $25$~nm, resulting in systems of up to $380$ million atoms with a total pillar height of $300$~nm (Fig.~\ref{fig:setup}a). These pillars are smaller than their experimental counterparts but have identical layer thickness and aspect ratio. The interaction between Cu and Au was modeled using a tailor-made embedded atom method potential~\cite{Gola:2018}. The flat, rigid indenter was obtained by freezing the structure of a Cu$_{50}$Zr$_{50}$ metallic glass obtained by melting a random solid solution at 2500K and quenching it down to 0K at a rate of $6$~K~ps$^{-1}$. A purely repulsive Lennard-Jones potential with interaction parameters $\epsilon_\textrm{Cu} = 0.4093$, $\sigma_\textrm{Cu} = 2.338$, $\epsilon_\textrm{Au} = 0.4251$, $\sigma_\textrm{Au} = 2.485$ acts between pillar and indenter~\cite{Halicioglu:1975:619}. Note that the disordered nature of the indenter introduces finite friction between indenter and pillar. We pressed the indenter onto the pillar by displacing it at a constant applied strain rate of $\dot{\varepsilon}_\textrm{app}=0.8\times 10^{8}\textrm{ s}^{-1}$. The whole pillar was kept at 300K using a using the Nos\'e-Hoover thermostat~\cite{Shinoda:2004:NPT_motion_equation} with a relaxation time constant of $0.5$~ps. A few rows of atoms at the bottom were fixed in space to anchor the pillar to the substrate.

We introduced different sources of defects in a controlled manner into our MD model: 1) Interface defects: Since Cu and Au are miscible, we intermix the interface between Cu and Au layers by randomly flipping Cu and Au atoms over a finite interface width of $15$\AA, such that the final concentration profile follows the error function predicted by simple Fickian diffusion. (See Ref.~\onlinecite{Gola2018-wb} for details.) 2) Surface defects: We introduced surface roughness on the pillar by cutting atoms above a plane that follows random self-affine scaling~\cite{Jacobs2017-ou,Persson2005-hx} with Hurst exponent $0.8$ and root-mean square (rms) slope of $0.1$ (Fig.~\ref{fig:setup}). 3) Bulk defects: As a representative volume defect, we introduced screw dislocations at random positions and orientations.

We quantified experiment by estimating the stress $\sigma$ inside the pillar before the nucleation of the first shear band (i.e. between the states shown in Figs.~\ref{fig:sem_exp}a and b). To do so, we extracted the cross section $d$ of the pillar at a position $1/5$ between top of the pillar and the Si substrate from SEM images such as those shown in Fig.~\ref{fig:sem_exp}. This gives a measure for the true strain in the pillar~\cite{schwaiger:2012:jmr}, $\varepsilon=\textrm{ln}\left(1+\left(d-d_0\right)/d\right)$ where $d_0$ is the initial diameter. Assuming rotational symmetry, it also gives an estimate of the cross sectional area, $A=\pi d^2/2$. The stress was then obtained from indenter force $F$ and area, $\sigma=F/A$.
Simulations are evaluated similarly. We computed the area $A$ from the convex hull of the cross section at the same position along the pillar (Fig.~\ref{fig:setup}c). Since experiments have only access to a side view and must assume rotational symmetry, we also computed the semi-minor and semi-major axes of the pillar and used their lengths to estimate the error in the determination of $A$ (see Fig.~\ref{fig:setup}c). Results obtained for different definitions of $A$ (smallest and largest cross section, exact convex hull) are indistinguishable from each other (see Supplementary Section S-II).

\begin{figure}
\centering
    \includegraphics[width=8.5cm,keepaspectratio]{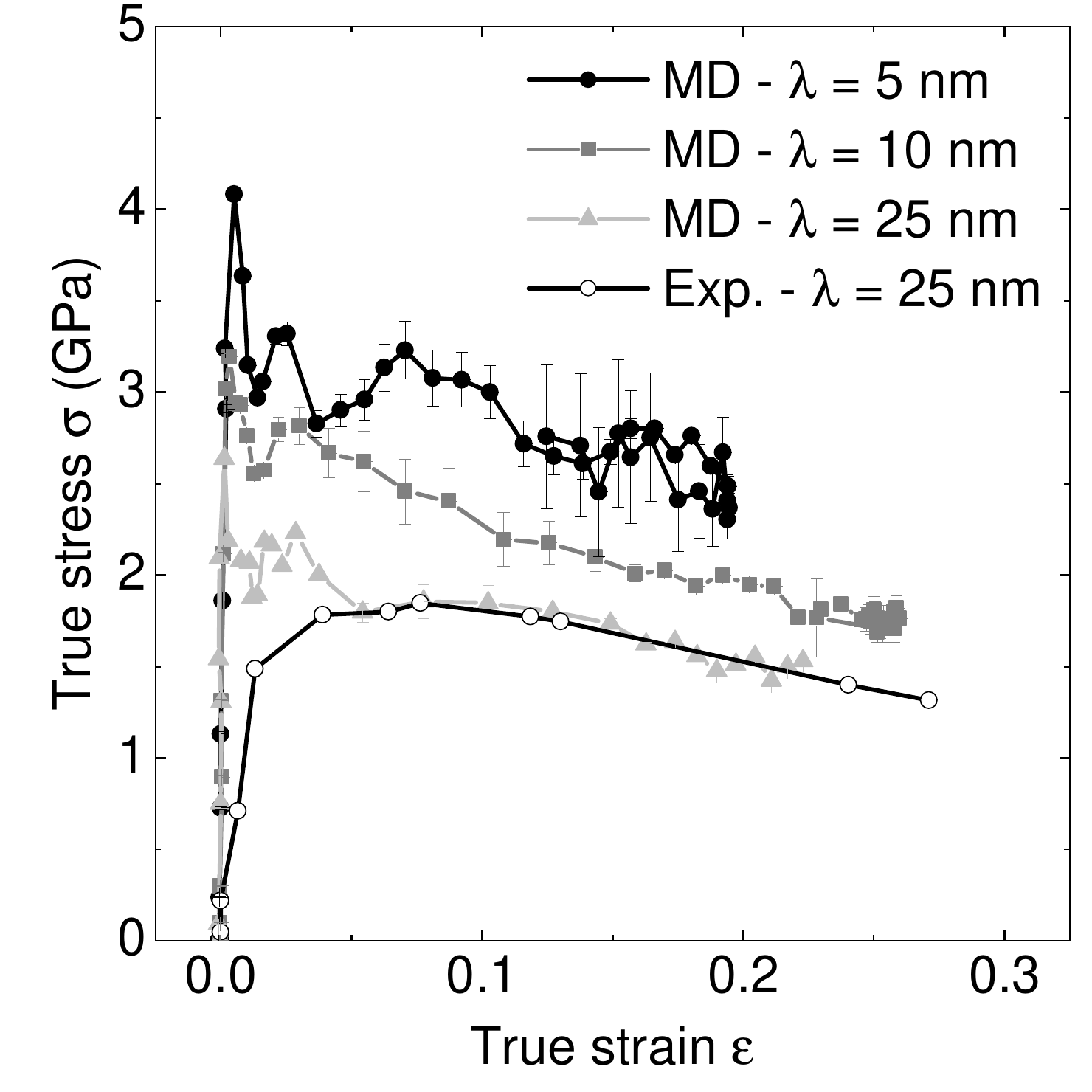}
    \caption{Stress-strain curves of pillar compression obtained in experiments and through MD simulations for different layer thickness $\lambda$. The lateral true strain and the area required to compute $\sigma$ are determined from reference cross sections at $1/5$ of the pillar height from the top of the pillar in all cases. The error bars of the simulated data are obtained by repeating the area measurement at distances $\pm 1$~nm of the reference cross section.}
    \label{fig:stress_strain}
\end{figure}

Experimental data is shown by the open symbols in Fig.~\ref{fig:stress_strain}. The stress rose to a maximum of $\sigma \sim 1.8$~GPa at $\varepsilon \sim 4\%$ strain and then dropped during subsequent deformation. This drop is not an indication of shear softening but arises because we do not use the contacting area, but the area $1/5$ from the pillar's top, to estimate the stress. Fig.~\ref{fig:stress_strain} also shows the simulated stress-strain curves for pillar models with interface and surface defects. The flow stress depends on layer thickness $\lambda$ and roughly decreases with $\lambda^{-1/2}$. Our calculation at experimental scales ($\lambda=25$~nm) reproduces the experimental curves in the flow region, given that we introduced at least surface roughness into our system.

\begin{figure}
\centering
    \includegraphics[width=8.5cm,keepaspectratio]{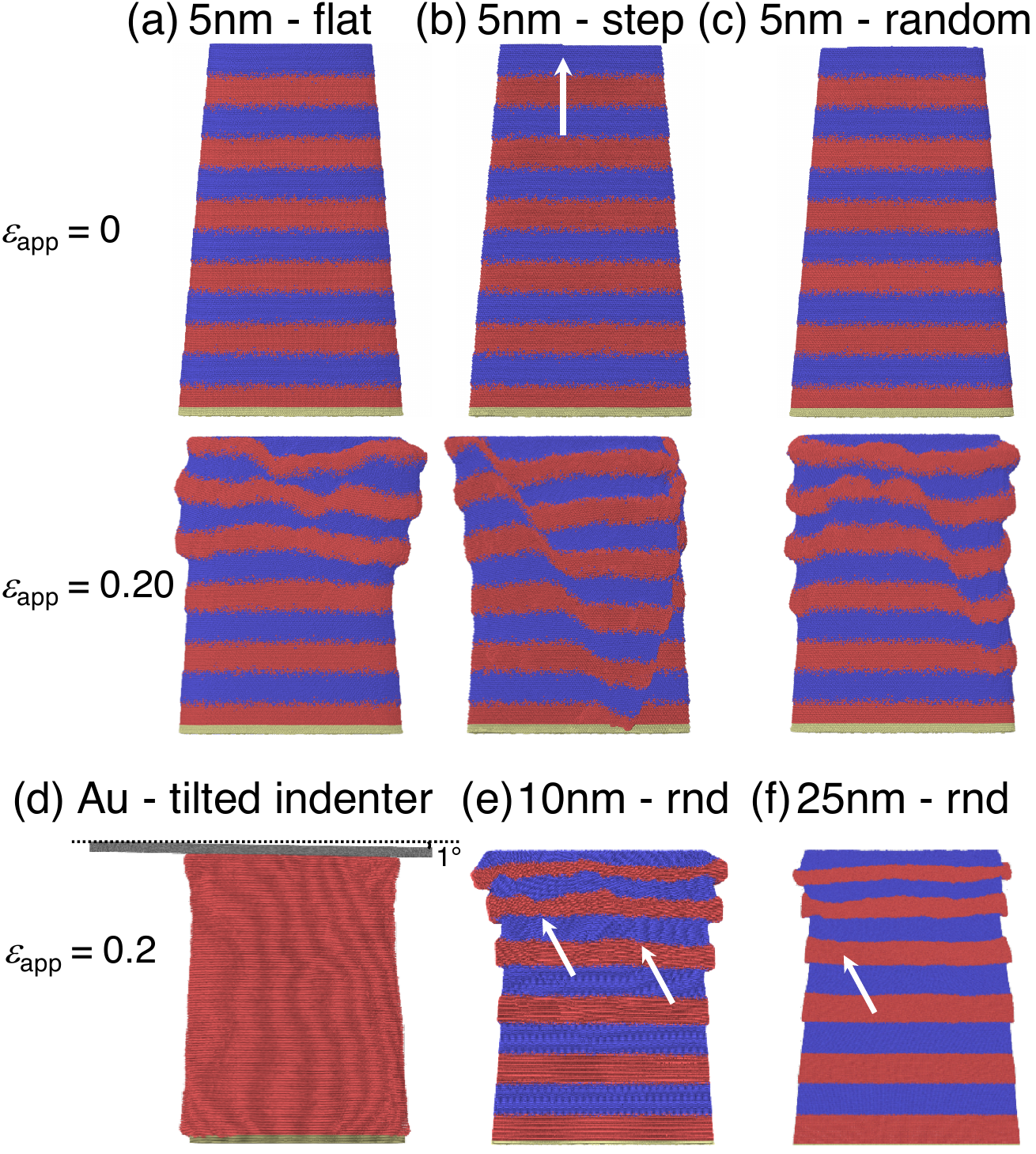}
    \caption{Comparison of deformation mechanism for (a) an atomically flat nanopillar, (b) nanopillars with an atomic step at the surface, (c, e, f) nanopillars with random roughness on the top with a root-mean-square slope of $0.1$. Panel (d) shows an Au nanopillar with a $1^{\circ}$ tilted indenter leading to a stress concentration at the pillar edge. The top row in (a-c) corresponds to the undeformed stage of the systems, while the bottom row corresponds to the systems after normal compressive strain of $0.2$. Layer thickness ranges from 5~nm (a-c), to 10~nm (e) and 25~nm (f). The indenter has been removed for clarity for the multilayer systems. Atoms are color coded after their type, Cu are in blue, Au are in red and fixed atoms are in yellow. Arrow in (b) show the initial position of step at the surface. Arrows in (e, f) show the location of initial formation of shear bands.}
    \label{fig:MD_failure}
\end{figure}

All stress-strain curves of Fig.~\ref{fig:stress_strain} show only the initial stages of deformation, before the first shear band nucleated in experiments (Fig.~\ref{fig:sem_exp}) or simulations. Further deformation in our simulations can be classified as occurring homogeneously (Fig.~\ref{fig:MD_failure}a and d) or heterogeneously through the formation of a shear band (Fig.~\ref{fig:MD_failure}b,c,e and f). 
Formation of a shear band eventually led to a failure-mode similar to the one observed experimentally (Fig.~\ref{fig:sem_exp}). A key observation in our simulations is that perfectly flat surfaces always lead to homogeneous deformation (Fig.~\ref{fig:MD_failure}a) while rough surfaces show heterogeneous deformation and failure (Fig.~\ref{fig:MD_failure}\textbf{}b,c,e and f).

To clarify the role played by roughness we created pillars with the simplest model for ``roughness'', a single atomic step on the surface (Fig.~\ref{fig:MD_failure}b). This model ``roughness'' already led to a deformation mechanism dramatically different from perfectly flat surfaces. A shear band is clearly visible already at an applied strain of $\varepsilon_\textrm{app}=0.20$, manifested by a series of kinks in the Cu$\vert$Au heterointerfaces and extrusion of a wedge-shaped part of the pillar (Fig.~\ref{fig:MD_failure}b, bottom row).

It is remarkable that the single step is sufficient to nucleate a shear-band. This nucleation occurs because edges concentrate stress~\cite{johnson_contact_1985} that trigger the emission of a single dislocation into the bulk. The dislocation leaves behind steps at the Cu$\vert$Au heterointerfaces, essentially imprinting the surface structure into the bulk of the material. Once a shear band has nucleated it will accommodate all subsequent deformation since the steps or kinks created by the band themselves concentrate stress if the elastic constants differ between the layers. Figure~\ref{fig:MD_failure}b also shows that the individual pillar can host more than one shear band. The final snapshot of this figure clearly shows an extruded, wedge-shaped region of the pillar that is bounded by two shear bands. The deformation of the pillar is strikingly similar to the experimental result shown in Fig.~1. We were able to nucleate shear bands from steps, rough surfaces, or tilted indenters that all lead to stress concentration somewhere on the surface of the pillar.

Our explanation for the formation of the shear band relies on the existence of domains with varying elastic modulus. We therefore carried out control calculations using single crystalline Au pillars. Those pillars deformed homogeneously even in the presence of surface steps, self-affine roughness or a tilted indenter (Figure~\ref{fig:MD_failure}d and Supplementary Section S-III). We observed that after a dislocation nucleated at the surfaces it subsequently traversed the full pillar, vanishing at the side walls of the pillar and leaving behind a complementary step. Unlike in nanolaminates, this dislocation does not imprint its signature into the bulk of the material. 
While the surface flaws are the reasons for the nucleation of an initial dislocation that constitutes the onset of the shear band, the existence of alternating sequences of hard and soft materials is the fundamental reason for its formation~\cite{Knorr2013-fq}.

In summary, we have obtained the strength of Cu$\vert$Au nanolaminate pillars from experiments and atomic-scale simulation that show excellent agreement. The strength of the pillars decreases roughly as the square root of the layer thickness down to the thinnest systems consisting of 5~nm thick layers. Our pillars localized shear in shear bands that led to catastrophic failure of the material. We show that the nucleation process is extremely sensitive to surface flaws but the formation of the shear band is a result of the imprinting of the surface flaws into the interface structure of the nanolaminate. Since stress concentrations in the bulk can only occur if there is a contrast between the nanolaminate layers, a possible route to suppress the shear banding instability could be the search for modulus-matched nanolaminates.

\begin{acknowledgments}
\emph{Acknowledgments.} We thank Peter Gumbsch for helpful discussion. This research was partially supported by the Helmholtz Association and the Chinese Academy of Sciences through a joint research group (HCJRG-217 and GJHZ1401). L.P. acknowledges funding through the Deutsche Forschungsgemeinschaft DFG (grant PA 2023/2). All molecular dynamics calculations were carried out with LAMMPS~\cite{LAMMPS}. ASE~\cite{ASE} and OVITO~\cite{ovito} was used for pre-processing, post-processing and visualization. Computations were carried out on NEMO (University of Freiburg, DFG grant INST 39/963-1), ForHLR II (Steinbuch Center for Computing at \mbox{Karlsruhe} Institute of Technology, project ``MULTILAYER'') and JUQUEEN (J\"ulich Supercomputing Center, project ``hka18'').
\end{acknowledgments}


\newpage

\setcounter{figure}{0}
\renewcommand{\thesection}{S-\Roman{section}}
\renewcommand{\thefigure}{S-\arabic{figure}}
\renewcommand{\thetable}{S-\arabic{table}}
\renewcommand{\theequation}{S-\arabic{equation}}


\begin{center}
\large\bf{ Supplementary Material for \\
 ``Surface flaws control strain localization in the deformation of Cu$\vert$Au nanolaminates'' }
\end{center}

\section{Interfacial shear strength}

\subsection{Experiment}

\begin{figure}
\centering
    \includegraphics[width=1\linewidth,keepaspectratio]{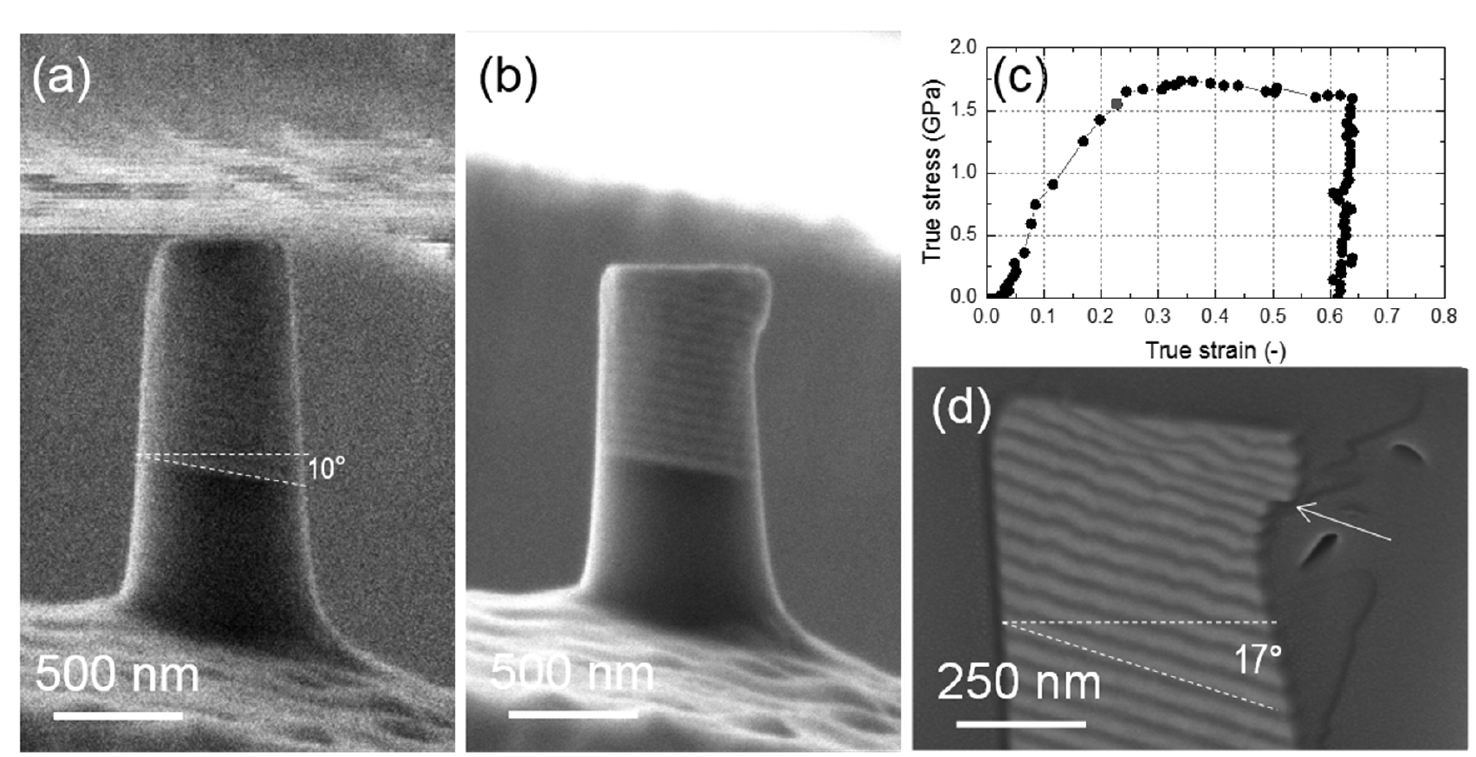}
    \caption{(a) SEM image of the $10^\circ$ tilted pillar prior to deformation. (b) SEM image of the $10^\circ$ tilted pillar after deformation up to $0.65$ strain. (c) True stress - True strain curve obtained for the deformation of the $10^\circ$ tilted pillar. (d) Post-mortem SEM cross section image of a $17^\circ$ tilted pillar. }
    \label{fig:S_sem_tilted_pillar}
\end{figure}

Pillars with interfaces tilted at an angle of $10^{\circ}$ and $17^{\circ}$ to the horizontal were prepared by FIB milling. They were deformed in situ in the scanning electron microscope to determine the true stress vs. true strain curves (FEI Nova NanoLab 200 and Nanomechanics InSem nanoindenter). The $10^{\circ}$ pillar was deformed to a strain of $\approx0.65$ at the pillar top, where the deformation localized (Fig.~\ref{fig:S_sem_tilted_pillar}). The deformation was stable and shear in the direction of the interface was not observed. At the maximum strain a shear stress acting along the interface of $\approx0.2$~GPa was observed. In case of the $17^{\circ}$ pillar, more pronounced steps on the pillar side-face were observed (marked by arrow in Fig.~\ref{fig:S_sem_tilted_pillar}d), while the pillar did not fail catastrophically. The maximum shear stress along the interface was $\approx0.3$~GPa.

\subsection{Simulation}

\begin{figure}
\centering
    \includegraphics[width=0.5\linewidth,keepaspectratio]{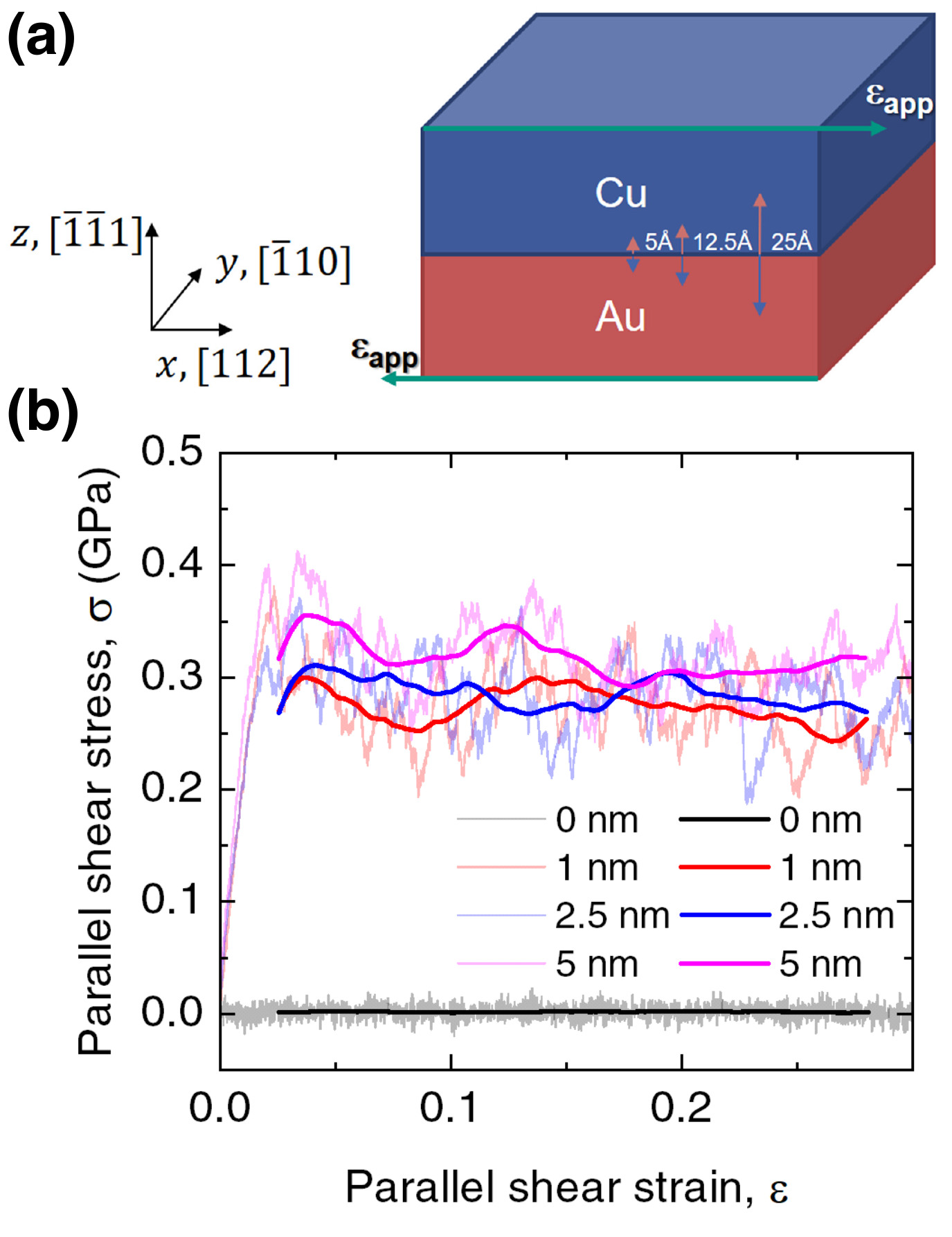}
    \caption{(a) Illustration of the Cu$\vert$Au nanolaminate simulation geometry and the applied strain direction used for the determination of the interfacial shear strength. (b) Stress-strain curves during simple shear at $300$~K parallel to the heterointerface of the Cu$\vert$Au nanolaminate system with a layer thickness of $5$~nm and intermixing width $w = 0, 1, 2.5$ and $5$~nm. Thick lines are moving averages over a strain interval $\pm0.025$ around the respective data point, thin lines show the full data.}
    \label{fig:S_stress-strain_shear}
\end{figure}

We used representative volume elements to compute the interfacial shear strength of the Cu$\vert$Au using molecular dynamic calculations. The system represented in Fig.~\ref{fig:S_stress-strain_shear}a was composed of a single bilayer with a layer thickness of 5~nm and periodic boundary conditions in all directions. The system was composed of approximately $200,000$ atoms with a box size of approximately $17\times19\times10$ nm$^3$ along the $x$, $y$, and $z$ directions, respectively.

Before straining, the systems was relaxed at $300$~K for $500$~ps using the Nos\'e-Hoover/Andersen~\cite{Shinoda:2004:NPT_motion_equation} ensemble without any strain. Simple shear strain was applied along $[112](1\bar{1}\bar{1})$ directions for shear parallel to the nanolaminate interfaces by homogeneously deforming the box. Our notation $[abc](hkl)$ for simple shear reports both the direction of shear $[abc]$ and the plane of shear $(hkl)$. We used a strain rate of $10^8$~s$^{-1}$ in all cases; strain rate dependence of stress is negligible at these rates in FCC metals~\cite{horstemeyer2001}.
For an atomically sharp interface, the nanolaminate responded to this deformation with a shear stress of a few MPa (Fig.~\ref{fig:S_stress-strain_shear}b). This is a clear sign that the system reacted by gliding along the heterointerface. As we probed more realistic systems with Cu and Au intermixed at the interface (see Ref.~\onlinecite{Gola2018-wb}), we observed that the yield stress increased in all the cases to approximately $0.3 - 0.35$~GPa. The intermixing width has only a small impact on the yield stress and interfacial shear strength which means that most of the strengthening comes from heterogeneities introduced close the heterointerface. Those act as pinning point for the interfacial dislocation network~\cite{Gola:2018}.

\section{Determination of the pillar cross-section}
\begin{figure}
\centering
    \includegraphics[width=1\linewidth,keepaspectratio]{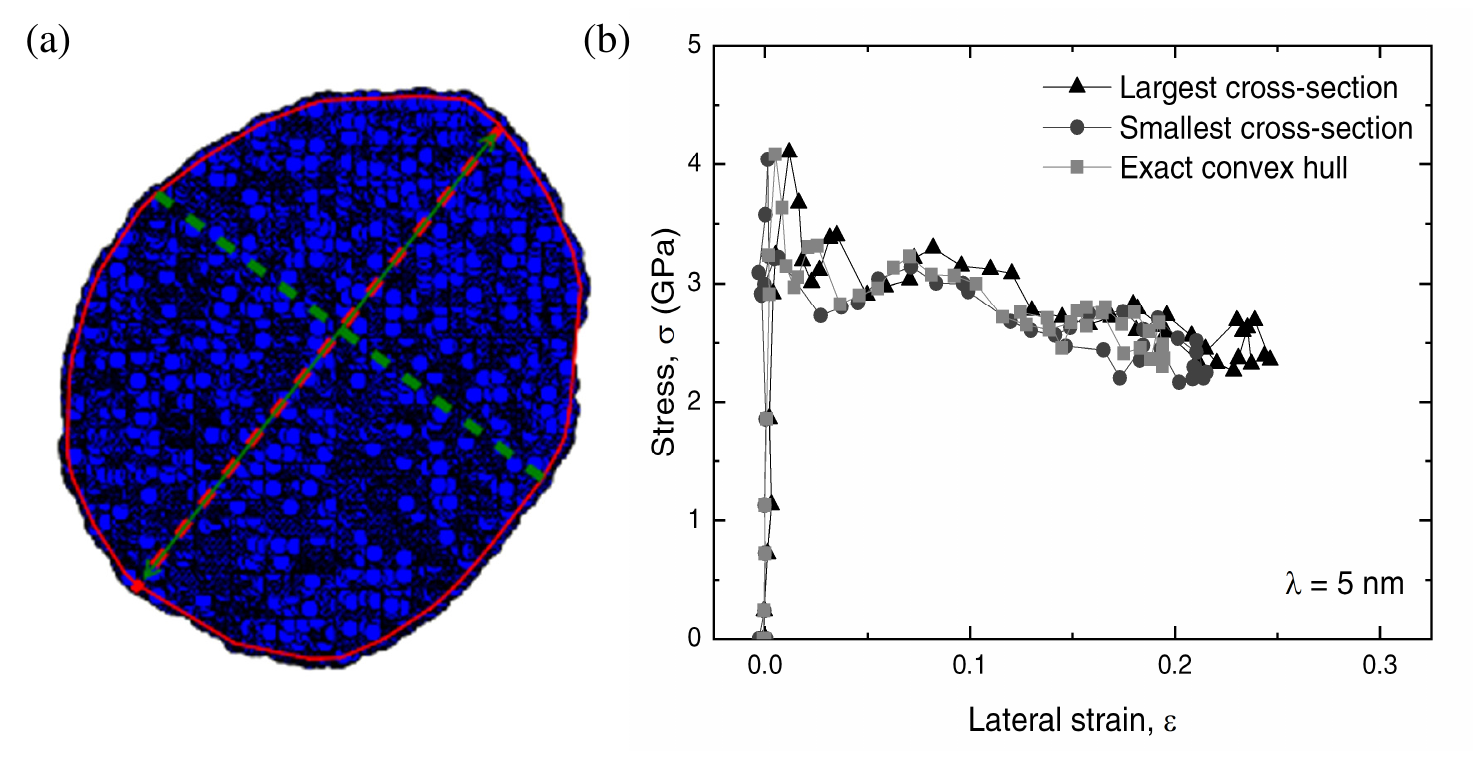}
    \caption{(a) Example of a cross section extracted from a pillar during indentation. The set of blue dots show the atom positions, the red solid line displays the convex hull of this given set of points. The red and green dashed lines represent the largest and smallest apparent diameters, respectively. (b) Stress-strain curves of pillar compression for the MD calculation with layer thickness $\lambda = 5$~nm.}
    \label{fig:S_cross_section}
\end{figure}

To facilitate comparison with experiments, we evaluated the MD simulations in the same way the experiments were evaluated: Stress $\sigma=F/A$ is computed by dividing the force $F$ on the indenter by the cross-sectional area $A$ at a position $1/5$ along the pillar from its top. Since experiments only have access to a side view (Fig.~\ref{fig:S_sem_tilted_pillar}) and must assume rotational symmetry, we investigated the influence of this assumption on the stress-strain curves. 
MD calculation allows us to access both the exact area and the apparent diameter of a given cross section as shown in Fig.~\ref{fig:S_sem_tilted_pillar}a. We computed the exact area $A$ from the convex hull of the cross section at the given height (red line in Fig.~\ref{fig:S_cross_section}a). We also computed the length of the semi-minor and semi-major axes of the pillar (as shown by the dashed lines in Fig.~\ref{fig:S_cross_section}a). With these measurements we determined the lateral strain in the pillar, 
$\varepsilon=\mathrm{ln} \left(1+\left(d-d_0\right)/d_0\right)$ where $d_0$ is the initial diameter.
Fig.~\ref{fig:S_cross_section}b shows the results obtained for the different definitions of the cross-sectional area $A$ (smallest and largest cross section, exact convex hull) for an exemplary calculation. We observe for all the cases a yield at $\sigma\approx4$~GPa  and $\varepsilon$ ranging from $0.1$\% to $1$\% followed by some strain softening. The maximal lateral strain is achieved for the largest cross-section definition with $\varepsilon\approx25$\%  the smallest cross section reaches $\varepsilon\approx22$\% and the exact convex hull area $\varepsilon\approx19$\%. In all the cases, the final stress value is around $\sigma\approx2.3$~GPa .
These results show that the assumption made in the experiments does not have a significant influence on the outcome of the stress-strain curves.

\section{Deformation of single-crystalline Au pillars}

\begin{figure}
\centering
    \includegraphics[width=1\linewidth,keepaspectratio]{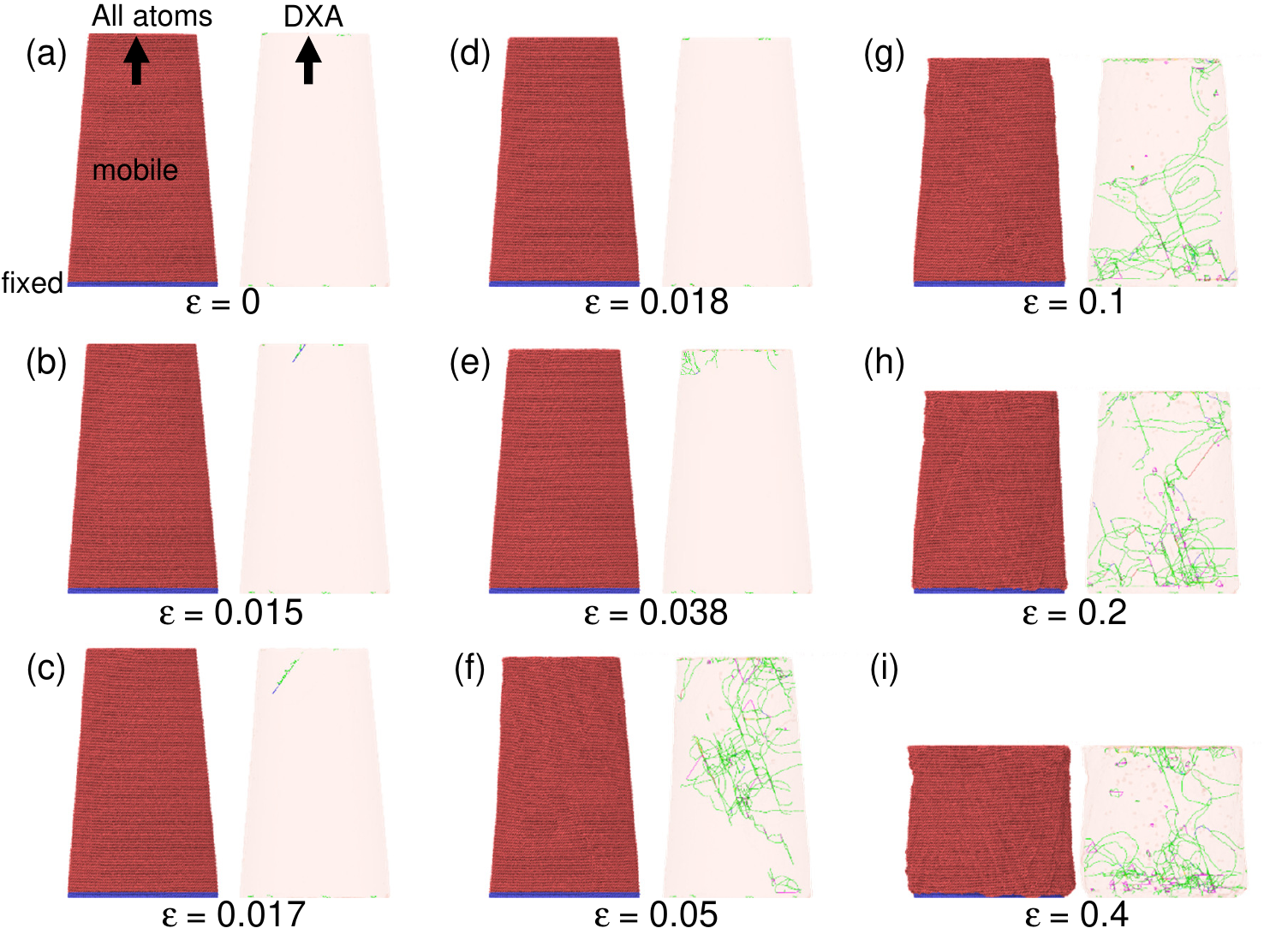}
    \caption{Deformation of an Au single crystal nanopillar of $60$~nm height under compression. The arrows in (a) mark the location of the single atomic step at the surface. Atoms are color coded after their mobility with red atoms being mobile and blue fixed. We used the dislocation extraction algorithm (DXA) to display the dislocations at each deformation stage. The shaded surface represents the nanopillar surface, Shockley partial dislocations are in green and stair-rod dislocations in purple.}
    \label{fig:S_Au_nanopillar}
\end{figure}

We carried out control calculations using single crystal Au pillars of $60$~nm height, equal to the total pillar height for the nanolaminate pillars with $\lambda = 5$~nm layer thickness. Fig.~\ref{fig:S_Au_nanopillar} show that the pillar deforms homogeneously even in the presence of a surface step. Alongside the atomic position we also show an analysis of the dislocation structure obtained with the dislocation extraction algorithm (DXA, Ref.~\cite{Stukowski:2012:DXA}). We obtain the same results for self-affine roughness (not shown here). 
We observed that after a dislocation nucleates at the surface (Fig.~\ref{fig:S_Au_nanopillar}b) it crosses the full pillar, vanishing at the sidewall and leaving behind a complementary step (Fig.~\ref{fig:S_Au_nanopillar}c-d). Unlike in nanolaminates, this dislocation does not imprint its signature into the bulk of the material. Further compression lead to new dislocations nucleating from the top pillar surface (Fig.~\ref{fig:S_Au_nanopillar}e). While some dislocations escape the pillar, others react in the bulk or pile up against the fixed layer at the bottom (Fig.~\ref{fig:S_Au_nanopillar}f-i).

\end{document}